\documentclass[graybox]{SNmult}

\usepackage{amsfonts}  
\usepackage{comment}

\newcommand{\chand}{\sqcap} 

\newcommand{\pand}{\wedge} 

\newcommand{\chor}{\sqcup} 

\newcommand{\por}{\vee} 

\newcommand{\blall}{\mbox{\large $\forall$}} 

\newcommand{\chall}{\hspace{1pt}\mbox{\Large $\sqcap$}} 

\newcommand{\blexists}{\mbox{\large $\exists$}} 

\newcommand{\chexists}{\hspace{1pt}\mbox{\Large $\sqcup$}} 

\newcommand{\pimplication}{\rightarrow} 

\newtheorem{theoremm}{Theorem}[section]

\begin{document} 
\title*{Do not throw out the baby: Clarithmetics as  alternatives to weak arithmetics}
\author{Giorgi Japaridze}

\institute{Giorgi Japaridze \at Department of Computing Sciences, Villanova University, 800 Lancaster Avenue, PA 19085, USA, \email{giorgi.japaridze@villanova.edu}}

\maketitle
 \abstract*{
Computability logic (CoL) provides a semantic foundation in which formulas represent interactive computational problems and validity corresponds to uniform algorithmic solvability.  
Building on this foundation, \emph{clarithmetics} — CoL-based axiomatic number theories — combine the full arithmetical strength of Peano arithmetic (PA) with explicit control over computational resources. In contrast to traditional bounded arithmetic and related complexity‑oriented systems, they strengthen rather than weaken PA. \\ This paper, after briefly surveying the relevant fragment of CoL,  presents the systems CLA4--CLA7 and  CLA11 of clarithmetic, and outlines their soundness and completeness with respect to natural classes of time, space, and so called amplitude complexities.
We argue  that, by weakening PA,  traditional  complexity-oriented systems of arithmetic “throw out the baby  with the bathwater,” discarding large amounts of innocent and useful arithmetical information and losing intensional flexibility essential for natural specification and program extraction. Clarithmetics avoid this loss while supporting direct extraction of optimal or near‑optimal algorithms from proofs and providing strong intensional completeness properties absent from bounded arithmetic and related systems. \\
A central message of the paper is that the argued advantages of clarithmetics deserve either acknowledgment or serious refutation from the weak‑arithmetics community. To date, neither has occurred.
} 

 \abstract{ 
Computability logic (CoL) provides a semantic foundation in which formulas represent interactive computational problems and validity corresponds to uniform algorithmic solvability.  
Building on this foundation, \emph{clarithmetics} — CoL-based axiomatic number theories — combine the full arithmetical strength of Peano arithmetic (PA) with explicit control over computational resources. In contrast to traditional bounded arithmetic and related complexity‑oriented systems, they strengthen rather than weaken PA. \\ This paper, after briefly surveying the relevant fragment of CoL,  presents the systems CLA4--CLA7 and  CLA11 of clarithmetic, and outlines their soundness and completeness with respect to natural classes of time, space, and so called amplitude complexities.
We argue  that, by weakening PA,  traditional  complexity-oriented systems of arithmetic “throw out the baby  with the bathwater,” discarding large amounts of innocent and useful arithmetical information and losing intensional flexibility essential for natural specification and program extraction. Clarithmetics avoid this loss while supporting direct extraction of optimal or near‑optimal algorithms from proofs and providing strong intensional completeness properties absent from bounded arithmetic and related systems. \\
A central message of the paper is that the argued advantages of clarithmetics deserve either acknowledgment or serious refutation from the weak‑arithmetics community. To date, neither has occurred.
} 
 
\section{Foreword}
This contribution is dedicated to my dear friend and mentor, Sergei Artemov. 

We first met in 1985 at the 4th Soviet–Finnish Symposium on Logic in Telavi. Sergei was already a rising star in provability logic—confident, insightful, unmistakably brilliant. I, by contrast, was a student still searching for my direction, unsure whether logic was truly my calling or merely one of several possibilities. That encounter profoundly influenced the trajectory of my life.

From the very beginning, Sergei became a guiding force for me. His remarkable ability to inspire, encourage, and instill confidence shaped my earliest steps in research. Without his steady belief in me, I might easily have wandered into a different career. Instead, what began as mentorship blossomed into a close and joyful collaboration, with our joint papers on first‑order provability logic among its fruits. Those years remain among the most formative and rewarding of my professional life.

Attending Sergei's weekly lectures at Moscow State University—together with Lev Beklemishev, Konstantin Ignatiev, Mati Pentus, Vladimir Shavrukov, and other bright young minds—was pure intellectual pleasure. Sergei had a way of making every session feel like an invitation to deeper understanding, a shared adventure in ideas. It was during those years that I developed my feel for, and love of, formal arithmetic—the central theme of Sergei’s research and lectures, and also a focus of the present paper.

By the turn of the century, our research paths began to diverge: Sergei went on to create and develop what became the celebrated {\em Justification Logic}, while I moved toward what I would later name {\em Computability Logic (CoL)}. Yet Sergei has always remained for me a mentor, a model, and a spiritual compass in the world of logic.

I wish Sergei many more years of creativity, discovery, and the joy of guiding new generations of students—students who, like me, will find their lives transformed by his wisdom, generosity, and extraordinary ability to bring out the best in those around him.

\section{Purpose and outline}
Several years ago, in a bulk email message titled ``Strong alternatives to weak arithmetics,'' I challenged the community of researchers working on weak arithmetics to take a deliberate look at my arguments, 
found in \cite{cla11a}, for basing complexity‑oriented versions of formal arithmetic on CoL rather than on classical logic—and to either show those arguments invalid or, otherwise, 
endorse the idea and approach. Neither response followed. 

The present article serves the sole purpose of making that challenge ``official,'' by presenting it in the form of 
a published paper rather than an email message, in the hope that this format will give it a better chance of being heard. This is also a paper explaining the key elements of the challenge 
in a compact way, rather than those elements  being scattered behind the forest of dauntingly long papers. The article is based on a 2025 oral presentation at the Computational Logic Seminar guided by Sergei Artemov at the City University of New York Graduate Center.

The structure of the rest of this paper is as follows.
Section~3 very briefly and informally (re)introduces the idea of CoL and surveys its fragment relevant to our purposes.
Section~4 does the same for the family of CoL‑based formal number theories dubbed {\em clarithmetics} (the prefix ``cl'' standing for ``computability logic'').
The final Section~5 compares clarithmetics with weak arithmetics—mainly bounded arithmetic—that share similar motivations, and highlights the main reasons why I believe clarithmetics, which strengthen rather than weaken Peano arithmetic, offer an appealing alternative.

\section{Computability logic in a nutshell}
Computability logic (CoL) is an ongoing project with no foreseeable end.
Only a relatively modest fragment of it will be surveyed here—namely, the fragment  relevant to the subsequent sections.
The full language of CoL is considerably richer, containing 30+ connectives and quantifiers of various sorts.

I characterize CoL as a formal theory of computability in the same sense in which classical logic is a formal theory of truth.
Indeed, let us compare the two logics.
Semantically, classical logic revolves around the concept of {\em truth}, a property of {\em statements}.
Its utility lies in providing systematic answers to questions such as “Is \(P\) (always) true?” or “Does the truth of \(P\) (always) follow from the truth of \(Q\)?”.
Computability logic replaces truth with {\em computability}.
The latter is a property of {\em computational problems}, and formulas now represent such problems rather than true/false statements.
Accordingly, the classical questions become:

- “Is \(P\) (always) computable?”

- “Does the computability of \(P\) follow from the computability of \(Q\)?” --- or, more generally,  ``Is $P$ algorithmically reducible to $Q$?''.

Moreover, whenever the answers are positive, CoL allows one to actually construct an algorithm for \(P\), or to extract such an algorithm from an algorithm for \(Q\).
This makes CoL not merely a descriptive logical framework but a genuine {\em problem‑solving tool}.

Under CoL's approach, classical statements naturally turn out to be nothing but  just special---``elementary''---cases of computational problems, and classical truth becomes a special case of computability.
This makes classical logic   a conservative fragment of CoL, obtained by restricting attention to the elementary part of its language.

\subsection{Games}
What, then, is computability?
As it is a property of computational problems, before defining computability we must clarify what  computational problems are.
According to Church \cite{church}, a computational problem is a function to be computed.
This view, however, is too narrow.
Most tasks performed by computers are interactive and cannot be reduced to functions.

CoL instead views computational problems as zero-sum {\em games} between a machine (denoted \(\top\)) and its environment (denoted \(\bot\)).
We are “fans” of the machine—even when, in practice, we ourselves act as its environment.
Interacting with a computer is one of the few situations where you do not want to win: if you defeat the machine, the machine has malfunctioned as a tool.

Computability is understood as (algorithmic) winnability by the machine.
For readability, this article refrains from formal definitions, though they are fully developed elsewhere; see \cite{ali} for a comprehensive reference.

Games can be visualized as trees, as in Figure~1.

\begin{center}
\begin{picture}(322,150)
\put(159,132){\circle{16}}
\put(155,128){$\bot$}
\put(151,132){\line(-4,-1){88}}
\put(95,122){{\tiny $\top$}{\footnotesize $\alpha$}}
\put(159,124){\line(0,-1){14}}
\put(160,116){{\tiny $\bot$}{\footnotesize $\beta$}}
\put(167,132){\line(4,-1){88}}
\put(212,122){{\tiny $\bot$}{\footnotesize $\gamma$}}
\put(62,102){\circle{16}}
\put(58,98){$\top$}
\put(55,98){\line(-5,-3){31}}
\put(26,88){{\tiny $\bot$}{\footnotesize $\beta$}}
\put(69,98){\line(5,-3){31}}
\put(88,88){{\tiny $\bot$}{\footnotesize $\gamma$}}
\put(159,102){\circle{16}}
\put(155,98){$\top$}
\put(159,94){\line(0,-1){14}}
\put(161,84){{\tiny $\top$}{\footnotesize $\alpha$}}
\put(256,102){\circle{16}}
\put(252,98){$\bot$}
\put(249,97){\line(-5,-3){31}}
\put(221,88){{\tiny $\top$}{\footnotesize $\alpha$}}
\put(256,94){\line(0,-1){14}}
\put(257,84){{\tiny $\top$}{\footnotesize $\beta$}}
\put(262,97){\line(5,-3){31}}
\put(279,88){{\tiny $\top$}{\footnotesize $\gamma$}}
\put(20,72){\circle{16}}
\put(16,68){$\top$}
\put(104,72){\circle{16}}
\put(100,68){$\bot$}
\put(102,64){\line(-2,-3){9}}
\put(85,56){{\tiny $\top$}{\footnotesize $\beta$}}
\put(107,64){\line(2,-3){9}}
\put(113,56){{\tiny $\top$}{\footnotesize $\gamma$}}
\put(159,72){\circle{16}}
\put(155,68){$\top$}
\put(214,72){\circle{16}}
\put(210,68){$\bot$}
\put(211,64){\line(-2,-3){9}}
\put(194,56){{\tiny $\top$}{\footnotesize $\beta$}}
\put(222,56){{\tiny $\top$}{\footnotesize $\gamma$}}
\put(217,64){\line(2,-3){9}}
\put(256,72){\circle{16}}
\put(252,68){$\top$}
\put(256,64){\line(0,-1){14}}
\put(257,56){{\tiny $\top$}{\footnotesize $\alpha$}}
\put(298,72){\circle{16}}
\put(294,68){$\bot$}
\put(298,64){\line(0,-1){14}}
\put(299,56){{\tiny $\top$}{\footnotesize $\alpha$}}
\put(92,42){\circle{16}}
\put(88,38){$\top$}
\put(116,42){\circle{16}}
\put(112,38){$\bot$}
\put(202,42){\circle{16}}
\put(198,38){$\top$}
\put(226,42){\circle{16}}
\put(222,38){$\bot$}
\put(256,42){\circle{16}}
\put(252,38){$\top$}
\put(298,42){\circle{16}}
\put(294,38){$\bot$}
\put(120,10){{\bf Figure 1:} A game}
\end{picture}
\end{center}

\noindent Here each edge represents a move, with the prefix \(\top\) or \(\bot\) indicating which player may legally make that move.
Correspondingly, each complete or incomplete branch represents a legal run—the sequence of edge labels along that branch.
Each node  represents a {\em position}---the finite run that takes us from the root to that node---and the label \(\top\) or \(\bot\) on a node indicates which player would win if the play ended at that position. For example, in the game of Figure 1, if \(\top\) makes move \(\alpha\), \(\bot\) replies with \(\gamma\), and no further moves occur, the resulting position is \(\bot\)-labeled; hence \(\bot\) wins the run/position \(\langle \top\alpha, \bot\gamma \rangle\).
There are no restrictions on the order of moves: in some positions either player may move.
In the root (empty position) of Figure~1, for instance, \(\bot\) has two legal moves (\(\beta\) and \(\gamma\)), while \(\top\) has one (\(\alpha\)).
Either player may move first, depending on who can or wants to act sooner.
This freedom sharply distinguishes CoL games from most games studied in logic, where players typically are expected to make moves in a strictly alternating order.
Traditional computational problems—functions to be computed—appear in CoL as very simple games of depth 2, illustrated in Figure~2.

\begin{center}
\begin{picture}(322,160)
\put(175,142){\circle{16}}
\put(-4,109){\scriptsize Input}
\put(171,138){$\top$}
\put(175,134){\line(-3,-1){106}}
\put(83,109){\tiny $\bot 0$}
\put(175,134){\line(0,-1){34}}
\put(163,109){\tiny $\bot 1$}
\put(175,134){\line(3,-1){106}}
\put(226,109){\tiny $\bot 2$}
\put(175,134){\line(6,-1){119}}
\put(285,109){\Huge ...}
\put(-4,59){\scriptsize Output}
\put(65,92){\circle{16}}
\put(61,88){$\bot$}
\put(30,59){\tiny $\top 0$}
\put(65,84){\line(-1,-3){11}}
\put(46,59){\tiny $\top 1$}
\put(65,84){\line(-1,-1){34}}
\put(62,59){\tiny $\top 2$}
\put(65,84){\line(1,-3){11}}
\put(77,59){\tiny $\top 3$}
\put(65,84){\line(1,-1){34}}
\put(95,59){\large ...}
\put(65,84){\line(3,-2){34}}
\put(29,42){\circle{16}}
\put(25,38){$\bot$}
\put(53,42){\circle{16}}
\put(49,38){$\top$}
\put(77,42){\circle{16}}
\put(73,38){$\bot$}
\put(101,42){\circle{16}}
\put(97,38){$\bot$}
\put(175,92){\circle{16}}
\put(171,88){$\bot$}
\put(140,59){\tiny $\top 0$}
\put(175,84){\line(-1,-3){11}}
\put(156,59){\tiny $\top 1$}
\put(175,84){\line(-1,-1){34}}
\put(172,59){\tiny $\top 2$}
\put(175,84){\line(1,-3){11}}
\put(187,59){\tiny $\top 3$}
\put(175,84){\line(1,-1){34}}
\put(205,59){\large ...}
\put(175,84){\line(3,-2){34}}
\put(139,42){\circle{16}}
\put(135,38){$\bot$}
\put(163,42){\circle{16}}
\put(159,38){$\bot$}
\put(187,42){\circle{16}}
\put(183,38){$\top$}
\put(211,42){\circle{16}}
\put(207,38){$\bot$}
\put(285,92){\circle{16}}
\put(281,88){$\bot$}
\put(250,59){\tiny $\top 0$}
\put(285,84){\line(-1,-3){11}}
\put(266,59){\tiny $\top 1$}
\put(285,84){\line(-1,-1){34}}
\put(282,59){\tiny $\top 2$}
\put(285,84){\line(1,-3){11}}
\put(297,59){\tiny $\top 3$}
\put(285,84){\line(1,-1){34}}
\put(315,59){\large ...}
\put(285,84){\line(3,-2){34}}
\put(249,42){\circle{16}}
\put(245,38){$\bot$}
\put(273,42){\circle{16}}
\put(269,38){$\bot$}
\put(297,42){\circle{16}}
\put(293,38){$\bot$}
\put(321,42){\circle{16}}
\put(317,38){$\top$}
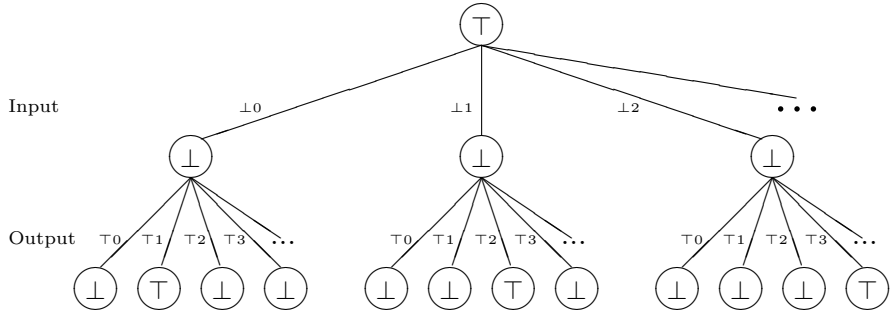
\put(65,10){{\bf Figure 2:} The successor function as a game}
\end{picture}
\end{center}

\noindent In such a game, the upper‑level edges represent possible inputs chosen by the environment, explaining their \(\bot\)-prefixes.
The lower‑level edges represent outputs generated by the machine, hence their \(\top\)-prefixes.
The root is \(\top\)-labeled because, with no input provided, the machine has nothing to answer for.
The middle nodes are \(\bot\)-labeled because an unanswered input means the machine loses.
Each group of bottom nodes contains exactly one \(\top\)-labeled node, reflecting the fact that a function has a unique correct value at each argument.

CoL sees no reason to restrict attention exclusively to such special games.
We may allow deeper or even infinite branches to model long‑running interactive tasks, and we may allow arbitrary patterns of \(\top\) and \(\bot\) labels.
For example, in computing \(1/x\), it is natural to label the node reached by input \(0\) with \(\top\), since the function is undefined there and the machine cannot be penalized for failing to produce an output.
We may also generalize in the opposite direction by allowing {\em elementary games}—games with no moves at all.
These correspond to classical propositions or predicates.
A true classical proposition (e.g., \(\top\) or “Snow is white”) is the elementary game automatically won by the machine; a false one (e.g., \(\bot\) or “Snow is black”) is automatically lost.
Thus classical logic is precisely the elementary fragment of CoL---the fragment that only allows games of depth $0$.

\subsection{Game operations} Below is  a very brief informal overview of the game operations used in Section 4 as logical operators of the language of clarithmetics. See \cite{ali} for strict definitions, intuitive insights and examples. 

{\em Choice conjunction \(\chand\):}
This operation forms a game in which the environment makes the first move, choosing between the left and right component.
If the environment fails to move, the machine wins by default.
Once the choice is made, play continues according to the chosen component.

{\em Choice disjunction \(\chor\):}
This is similar to choice conjunction, except that the machine makes the initial choice and loses if it fails to do so.

{\em Choice quantifiers \(\chall\) and \(\chexists\):}
Assuming the universe of discourse is \(\mathbb{N}\), the choice universal quantification \(\chall x\,G(x)\) is the infinite choice conjunction \(G(0)\chand G(1)\chand\ldots\), and the choice existential quantification \(\chexists x\,G(x)\) is the infinite choice disjunction \(G(0)\chor G(1)\chor\ldots\).
These allow us---instead of drawing trees---to compactly express various computational problems. For instance, the problem of computing the successor function shown as a tree in Figure 2 can now be simply written as \(\chall x\chexists y\,(y=x+1)\).

{\em Negation \(\neg\):}
Negation is role‑switching: \(\neg G\) is obtained from \(G\) by interchanging the roles of \(\top\) and \(\bot\).
For example, if \(G\) is the game of chess from White’s perspective, then \(\neg G\) is the same game from Black’s perspective.

{\em Parallel conjunction \(\pand\) and parallel disjunction \(\por\):}
These represent simultaneous play of two games \(A\) and \(B\).
In order to win \(A \pand B\), the machine must win both components; in \(A \por B\), winning either component suffices.

{\em Parallel implication \(\pimplication\):}
Defined as \(\neg A \por B\), this operation captures a form of reduction: the machine may treat \(A\) as a resource (because it appears negated) while attempting to win \(B\).

{\em Blind quantifiers \(\blall\) and \(\blexists\):}
\(\blall x\,G(x)\) differs from \(\chall x\,G(x)\) in that no value is specified for the quantified variable by either player.
In order to win, the machine must play in a way that guarantees a win in \(G(x)\) regardless of the value of \(x\). As always, \(\blexists\) is dual to \(\blall\), and \(\blexists x\,G(x)\) can be defined as \(\neg \blall x\,\neg G(x)\).

Classical operators—those written with their usual symbols—are conservative generalizations of their classical counterparts: when restricted to elementary games, they behave exactly as in classical logic.

See \cite{ali,Japfin} for explanations of how CoL relates to linear logic \cite{Gir87}, or to its game-semantical precursors \cite{Lor59,Hintikka73,Bla92}, with Blass's \cite{Bla92} game semantics for linear logic being the most immediate one.

\section{Clarithmetics}
Here we  briefly survey several systems of clarithmetic. See \cite{cla4,cla5,cla11a,ali} for full details and additional insights.

Clarithmetics are axiomatic number theories based on CoL, analogous to Peano arithmetic (PA) and the various bounded arithmetics based on classical logic. 
The language contains \(0\), successor (\('\)), addition \(+\), multiplication \(\times\), equality \(=\); the logical layer syntactically augments the standard vocabulary 
\(\neg,\pand,\por,\pimplication,\blall,\blexists\)---conservatively (re)interpreted as explained in Section~3.2---with choice operators \(\chand,\chor,\chall,\chexists\). 

Nonlogical axioms include the standard PA axioms, with the induction axiom scheme limited to \(\chand,\chor,\chall,\chexists\)-free formulas among them. 
Extra-Peano axioms are typically limited to the single \(\chall x\chexists y\,(y=x+1)\) (i.e., \(\chall x\chexists y\,(y=x')\)) asserting the computability of the successor function; some systems may require just a couple of equally 
simple additional extra-Peano axioms. On top of all that, some sort of a  ``{\em constructive  induction}'' rule is usually present. The logic running in the background is CoL rather than classical logic.

Let \(|x|\) denote the standard pseudoterm for the length of the binary numeral for \(x\). Boundedness is a syntactic discipline on the placement of choice quantifiers: a formula is {\em exponentially bounded} if every subformula \(\chall x\,E\) or \(\chexists x\,E\) of it occurs only in the  guarded form \( \chall z\,(|z| \le t \pimplication E)\) or \(\chexists z\,(|z| \le t \pand E)\) where \(t\) is a \(\{0,',+,\times\}\)-term built from variables other than \(z\). {\em Polynomially bounded} formulas are defined similarly but with the variables of \(t\) occurring between size bars just like $z$ does in $|z|$.

The systems CLA4 through CLA7 share the same language and the same axioms: all axioms of PA, together with the previously  mentioned additional extra-Peano axiom \(\chall x\chexists y\,(y=x+1)\).\footnote{Well, CLA4 in fact additionally requires the axiom \(\chall x\chexists y\,(y=2x)\), but overlooking this fact for the sake of simplicity is hopefully forgivable in this informal overview.}
Due to the intrinsic computational  content and strength ``hidden'' in CoL, this seemingly innocuous axiom turns out to be sufficient to derive all computability‑theoretic facts expressible in the system.

The four systems  only differ from each other in the constructive induction rule that is present along with Peano induction axioms. In CLA4, this rule takes the form
\[
\frac{A(0), \ \ \ \ \ A(x)\pimplication A(2x), \ \ \ \ \ \ A(x)\pimplication A(2x+1)}{\chall x\,A(x)},
\]
where \(A(x)\) is polynomially bounded. The constructive induction rule of CLA5--CLA7 is
\[
\frac{A(0), \ \ \ \ \ A(x)\pimplication A(x+1)}{\chall x\,A(x)},
\]
where \(A(x)\) is polynomially bounded in the case of CLA5,  exponentially  bounded in the case of CLA6, and has no restrictions at all in the case of CLA7.

With the standard notions of time and space complexity naturally and conservatively generalized to all interactive problems expressible in the language of clarithmetic, the main theorem is:
\begin{theoremm}
Under the standard arithmetical interpretation, and where +TA in clause 3 means adding to the system all true sentences of the language of Peano arithmetic as additional axioms, we have:
\begin{enumerate}
  \item {\bf Soundness:} Every theorem of CLA4 represents (expresses) an arithmetical problem with a polynomial time solution. Such a solution can be automatically extracted from a proof of the theorem.
  \item {\bf Extensional completeness:} Every arithmetical problem with a polynomial time solution is represented by some theorem of CLA4.
  \item {\bf Intensional completeness:} Every formula representing an arithmetical problem with a polynomial time solution is a theorem of CLA4+TA.
  \item Similarly for CLA5, but with polynomial space instead of polynomial time.
  \item Similarly for CLA6, but with elementary recursive time (=space).
  \item Similarly for CLA7, but with primitive recursive time (=space).
\end{enumerate}
\end{theoremm}

\medskip
\noindent\textbf{Examples.} Let us just consider CLA4, with its soundness and completeness with respect to polynomial time computability in mind.
\begin{itemize}
 \item The formula \(\chall x\chall y\chexists z\,(z=x\times y)\), expressing the problem of computing $\times$,  is provable. Hence multiplication is polynomial‑time computable.
 \item The formula \(\chall x\bigl(\blexists y_{>1}\blexists z_{>1}(x=y\times z)\chor \neg \blexists y_{>1}\blexists z_{>1}(x=y\times z)\bigr)\), expressing the problem of deciding primality, is provable. This reflects the fact---proven in \cite{agrawal}---that primality is polynomial‑time decidable.
 \item Consider \(\chall x\bigl(\blexists y_{>1}\blexists z_{>1}(x=y\times z)\pimplication \chexists y_{>1}\chexists z_{>1}(x=y\times z)\bigr)\),
 where the
 antecedent asserts the existence of nontrivial factors of \(x\) and the consequent demands that \(\top\) actually produce them. This formula is not known to be provable; if it were, polynomial‑time factoring and collapse of modern cryptography would follow.
 \item For any predicates  \(p\) and \(q\), the provability of \(\chall x\chexists y \bigl(p(x)\leftrightarrow q(y)\bigr)\) can be seen to mean that \(p\) is polynomial‑time mapping (many-one) reducible to \(q\).
 \item The provability of \(\chall x\bigl(q(x)\chor\neg q(x)\bigr)\pimplication \chall x\bigl(p(x)\chor\neg p(x)\bigr)\) can be seen to mean that
\(p\) is single‑query polynomial‑time Turing reducible to \(q\).
\end{itemize}

Finally, a few words on the author's personal favorite: the ``tunable'' clarithmetic CLA11, studied in \cite{cla11a,cla11b}. It has four simple extra-Peano  axioms  and two extralogical rules: constructive induction and comprehension. But rather than being a single specific  theory of clarithmetic, CLA11---more adequately written as CLA11($T,S,A$)---is in fact a scheme of  such  theories,
 with its three parameters \(T\), \(S\), and \(A\) governing simultaneous time, space, and so-called amplitude complexity. The latter is a novel complexity 
 measure indispensable in the context of interactive \ computations, concerned with the sizes of \(\top\)'s moves relative to the sizes of \(\bot\)'s moves. The three parameters are nothing but sets of terms used as bounds in the induction and comprehension rules of the system.
By choosing them appropriately, one obtains a theory sound and complete (in the sense established in Theorem~1 for CLA4-CLA7) with respect to any ``reasonable'' combinations of simultaneous time, space and amplitude complexities.
For example, the instance of CLA11 for polynomial time + polylogarithmic space + linear amplitude is obtained by mechanically setting:
\begin{itemize}
 \item \(T\) to  $(0,',+,\times)$-combinations of variables in the form $|x|, |y|, \ldots$;
 \item \(S\) to  $(0,',+,\times)$-combinations of variables in the form $||x||, ||y||, \ldots$;
 \item \(A\) to  $(0,',+)$-combinations of variables in the form \(| x|, |y|, \ldots\).
\end{itemize}

\section{Clarithmetics vs.\ mainstream weak theories}
We now turn to a comparison between clarithmetics and the traditional weak arithmetics such as Buss's \cite{Buss} bounded arithmetic and numerous other systems \cite{bbb2,cook,parikh,paris1,Sch,zambella} in a similar style.

\subsection{Proofs as programs}
One of the main motivations behind weak arithmetics is the ``proofs as programs'' paradigm.
The following quotation from \cite{Sch} captures the idea:
\begin{quote}
It is well known that it is undecidable in general whether a given program meets its specification.
In contrast, it can be checked easily by a machine whether a formal proof is correct.
From a constructive proof one can automatically extract the corresponding program, which, by its very construction, is correct as well.
This opens a way to produce correct software, for instance for safety‑critical applications.
Moreover, programs obtained from proofs are commented in an extreme sense, making them easy to apply, maintain, and adapt.
\end{quote}
This motivation applies equally well—and, I would argue, more naturally—to clarithmetics. Under the semantics-based approach of CoL,
algorithms extracted from proofs have optimal or near‑optimal complexities; one does not obtain the monstrous polynomials that sometimes arise in proof‑theory-based approaches.

Unlike clarithmetics, weak arithmetics attempt to use the classical language of arithmetic—designed for speaking about truth—to express computability.
This is like hammering a nail with a wrench: possible, but not natural.
CoL, by contrast, was designed from the ground up to express computational tasks.

The price paid by weak arithmetics is a loss of \emph{intensional} strength.
One can no longer freely write specifications in the natural language of arithmetic; many harmless and useful formulas become unwritable or unprovable.
They throw out the baby with the bathwater, that is. To compensate and bring some of the ``baby'' back, weak arithmetics typically have to introduce numerous new primitives and axioms. 
Clarithmetics, in contrast, strengthen rather than weaken Peano arithmetic, thus fully preserving the implicit information contained in the latter: information that is only limited due to the  inescapable Gödel  incompleteness theorems.

For applications, it is of course intensional rather than extensional strength that really matters. This is where clarithmetics with their intensional completeness shine, while their alternatives achieve only extensional completeness and, otherwise, are by definition intensionally incomplete and weak.
Suppose one wants a program computing a certain function.
In clarithmetics, one simply writes the specification in the natural way.
Intensional completeness guarantees that if \emph{any} equivalent specification is provable, then the one written is provable as well.
Extensional completeness, by contrast, guarantees only that \emph{some} equivalent formula is provable.
But the system may not know that the two formulas are equivalent, and the user may not know how to rewrite the specification into a provable form.
Weak arithmetics therefore require heavy preprocessing: one must rewrite the specification into a special syntactic form (e.g., a \(\Sigma^b_1\) formula) before it becomes provable.
But performing this preprocessing essentially amounts to solving the problem in advance.

\subsection{Separating complexity classes}
A second motivation for weak arithmetics is the hope that they may help separate complexity classes.
As Cook and Nguyen \cite{cook} put it, it ought to be easier to separate theories corresponding to complexity classes than to separate the classes themselves.
Here clarithmetics again have an advantage.
Separating theories \emph{intensionally} is easier than separating them \emph{extensionally}.
To separate two intensionally complete theories, it suffices to find a single formula provable in one but not the other.
To separate them extensionally, one must show that no formula equivalent to a given one is provable in the other theory—a far more difficult task.

\subsection{Further advantages}
Because clarithmetics fully preserve PA, reasoning within them is straightforward.
One may rely on ordinary arithmetical intuition without pretending that, for instance, exponentiation is not total or that other familiar facts fail.
In weak arithmetics, by contrast, one must work in an ``alternate reality'' where many basic arithmetical facts are no longer available.
Worse, each weak system corresponds to a different alternate reality, so expertise in one does not transfer to another.

As already pointed out, in weakening PA, traditional bounded arithmetics  discard large amounts of innocent and useful arithmetical information, along with the expressive flexibility that makes ordinary reasoning in arithmetic natural and transparent.
Clarithmetics require no changes to the non‑logical vocabulary of PA.
The systems CLA4–CLA7 use exactly the same function symbols as PA, plus a single extra‑PA axiom expressing the computability of the successor function.
CLA11 adds only a few more axioms.
Bounded arithmetic, by contrast, must introduce several new primitives (e.g., a primitive for \(\le\)) and dozens of axioms—thirty‑two in one standard presentation \cite{Buss}.
These primitives and axioms are needed only because the original language of arithmetic is no longer freely usable.

A number of approaches to weak arithmetics (cf. \cite{cook}) achieve completeness by adding axioms that arithmetize some nontrivial theory of computation, such as axioms representing complete problems for a complexity class.
These axioms become enormous and number‑theoretically opaque, as they encode graphs, computations, and other non‑arithmetical structures.
Clarithmetics, by contrast, as we saw, use extremely simple axioms, thus being more amenable to meta\-analysis.

Yet another major advantage of clarithmetics is that they naturally express interactive computational problems.
Bounded arithmetics do not.
Clarithmetics can express not only traditional function‑computing and predicate‑deciding tasks, but also arbitrary interactive tasks and relations between them—many with no established names in the literature.

It should also be noted that clarithmetics, unlike other arithmetics including PA, are categorical: they prove the decidability of addition and multiplication, and by Tennenbaum’s theorem there is only one model.
Whether this categoricity should be viewed as an advantage or an obstacle remains to be seen. It  is not yet clear how to reinterpret constructive (choice) operators to neutralize Tennenbaum’s theorem, which would be necessary to do if attempting to separate theories using   nonstandard models of arithmetic.

\end{document}